# Multistep Model for Predicting Upper-Limb 3D Isometric Force Application from Pre-Movement Electrocorticographic Features*

Jing Wu, *Student Member, IEEE*, Benjamin R. Shuman, Bingni W. Brunton, Katherine M. Steele, Jared D. Olson, Rajesh P.N. Rao, *Member, IEEE*, and Jeffrey G. Ojemann

*Abstract*—Neural correlates of movement planning onset and direction may be present in human electrocorticography in the signal dynamics of both motor and non-motor cortical regions. We use a three-stage model of jPCA reduced-rank hidden Markov model (jPCA-RR-HMM), regularized shrunken-centroid discriminant analysis (RDA), and LASSO regression to extract direction-sensitive planning information and movement onset in an upper-limb 3D isometric force task in a human subject. This mode achieves a relatively high true positive force-onset prediction rate of 60% within 250ms, and an above-chance 36% accuracy (17% chance) in predicting one of six planned 3D directions of isometric force using pre-movement signals. We also find direction-distinguishing information up to 400ms before force onset in the pre-movement signals, captured by electrodes placed over the limb-ipsilateral dorsal premotor regions. This approach can contribute to more accurate decoding of higher-level movement goals, at earlier timescales, and inform sensor placement. Our results also contribute to further understanding of the spatiotemporal features of human motor planning.

## I. INTRODUCTION

Decoding human arm movement has been a major goal of motor brain-computer interfaces (BCIs) for the development upper-limb neuroprostheses. Many previous decoding approaches have related recorded sensorimotor signals with specific virtual or robotic movements [1]–[3], allowing a potential BCI to reconstruct intended movements from a combination of basic signal changes. For example, primary motor cortex signals during individual finger movements could be combined to reconstruct multi-axis virtual cursor movements [4], [5], or pinch or grasp movements [6]–[8].

Alternatively, more abstract representations of movements, or movement plans, may be present in recorded brain signals prior to the onset of movement. The presence of these signals may especially be true in areas outside of the sensorimotor cortex. For instance, studies in non-human primates and human neuroimaging have established that, during an upper-extremity movement in a reach task [9]–[11], the parietal and premotor cortices play a key role in the representation of target trajectories and executing planned movements in the desired spatial coordinates.

An effective control signal for a BCI must maximize the information available, including preparation prior to the onset of movement. Patients with indwelling electrocorticography (ECoG) electrodes implanted for medically intractable epilepsy provide recordings with centimeter resolution combined with millisecond sampling over a large brain region. These signals offer an opportunity to detect correlates of the entire movement preparation and execution process. Previous work have demonstrated the capacity of ECoG signals to resolve the timing and interaction between regions during finger movements [8], upper-limb movements [1], [2], as well as premotor responses that precede the sensory and motor cortex activation up to 100ms in advance [12]. One previous work has also shown some success in detecting ECoG representations of onset and direction of hand-arm movements [13].

However, the temporal and spatial distributions of the recorded signal in motor preparatory regions are not well-established in human ECoG. It is yet unclear which non-motor regions contain ECoG signals with direction-distinguishing preparatory information, whether these signals are localized to just the contralateral hemisphere or both hemispheres, and how they are localized in time. It is also unknown whether correlates of preparatory activity are limited to reach trajectories, or also can generalize to other directional tasks such as isometric force application towards planned directions.

These unknown factors are particularly relevant to the engineering of BCI, as signals conveying information about the planning of a movement may allow for better resolution of differing task goals, inform locations for sensor placement, and enable decoding of these goals earlier than signals in the sensorimotor cortex.

In this work, we investigated the recorded ECoG activities of a patient with fortuitous bilateral implantation of electrodes in the parietal and arm-associated sensorimotor cortices. The patient performed a right arm isometric force task toward designated directions. We constructed a three-stage model to examine the localization and timing of direction-discriminable information, allowing for the parallel prediction of isometric force onset and direction.

*Research supported by grants from NSF EEC-1028725, NINDS 5R01NS065186, NIH 2K12HD001097, NIH 5U10NS086525, and the WRF Fund for Innovation in Neuroengineering.

J. Wu is with the department of Bioengineering, University of Washington, Seattle, WA 98105 USA. (phone: 858-729-8669; e-mail: jiwu@uw.edu).
B.R. Shuman is with the department of Mechanical Engineering, University of Washington, Seattle, WA 98105 USA (e-mail: brshuman@uw.edu)
B.W. Brunton is with the department of Biology, University of Washington, Seattle, WA 98105 USA. (e-mail: bbrunton@uw.edu).
K.M. Steele is with the department of Mechanical Engineering, University of Washington, Seattle, WA 98105 USA (email: kmsteele@uw.edu)
J.D. Olson is with the department of Rehabilitation Medicine, University of Washington, Seattle, WA 98105 USA. (email: jaredol@uw.edu)
R.P.N. Rao is with the department of Computer Science, University of Washington, Seattle, WA 98105 USA. (e-mail: rao@cs.washington.edu).
J.G. Ojemann is with the department of Neurosurgery, University of Washington, Seattle, WA 98105 USA. (e-mail: jojemann@uw.edu).

## II. METHODS

### A. Experimental Task and Data Acquisition

The patient is a 26-year-old right-handed male with intractable seizures thought to be related to a medial parietal lesion; invasive monitoring of bilateral parietal regions was recommended. The subject received pre-operative anatomical MRI, and was implanted with two frontoparietal 2×8 platinum electrode grids and two 1×8 strips (Ad-Tech, Racine, WA, USA), with 3mm electrode-diameters embedded in silastic with center-to-center distances of 10mm. Subsequently, X-ray CT imaging captured the location of the electrodes. These electrodes were sampled at 1220.7 Hz on a 128-channel Tucker-Davis Technology (Alachua, FL, USA) System 3 RZ5D Neurophysiology Base Processor with associated PZ5 NeuroDigitizer. The subject had normal motor function pre-operatively and throughout the experiments and inpatient stay. The patient gave written, informed consent through a protocol approved by the University of Washington Institutional Review Board.

The patient was comfortably positioned in front of one screen during the task, and was visually cued to apply isometric force in one of 6 directions (up, down, left, right, forward, back) to an affixed AMTI force transducer handle (Watertown, MA, USA) held in the right hand. Each trial consisted of a hold with approximately 15 lbs (67 N) of force and 2 seconds in duration, with an inter-trial interval of 3 seconds. 10 trials of each of 6 directions were recorded for a total of 60 isometric force trials.

### B. Data Preprocessing and Feature Extraction

Pre-operative MRI was co-registered with postoperative CT using Statistical Parametric Mapping (http://www.fil.ion.ucl.ac.uk/spm), with the pial surface reconstructed with FreeSurfer (http://freesurfer.net) and custom mapping and projection code [14] implemented in MATLAB (Natick, MA, USA). The two frontoparietal hemisphere 2×8 grid recordings were visually inspected for data quality, and channels with high level of artifacts and low impedance were rejected (1 in the contralateral L grid and 2 in R). The remaining 29 recorded channels were filtered with a common average reference filter $s_i' = s_i - J^{-1}\sum_j s_j$ for $J$ non-rejected channels within each separate lateral grid. The filtered signals were then analyzed for time-frequency content by continuous wavelet transform with the non-analytic Morlet wavelet defined by $\Psi(s\omega) = \pi^{-1/4} e^{[-(s\omega-\omega_0)^2]/2}$, where $\omega_0 = 6$ and $s$ reflects log scaling factors to obtain ⅛-octave resolution across pseudo-frequencies 2−200 Hz. We used the absolute magnitude of the wavelet coefficient time series at each scaling, and also extracted the lowpassed ECoG local motor potential (LMP) for each channel as outlined in [15] as an additional time series. All time series were binned by 50ms and normalized to unit variance.

Recorded force transducer voltages were converted to force using manufacturer sensitivity calibration data; Cartesian forces were transformed to spherical coordinates to obtain task force direction and magnitude. Subsequently, the initiation and end of recorded isometric force in each trial were carefully hand-labeled (examples shown with dotted force traces and vertical onset lines in Figure 3).

### C. Pre-Movement Classification of Force Direction

*jPCA-RR-HMM analysis:* We constructed a reduced-rank hidden Markov model (RR-HMM) [16] from the normalized binned time series using jPCA, a dimensionality reduction method designed to find orthonormal bases that also exhibit a rotational structure, with the expectation of a rhythmic repeating behavior from the signal dynamics [17]. To find these repeating mean trajectory templates, filtered time series were split into 2-second segments center-aligned to the initiation of movement, reflecting a time period of [−1000ms…+1000ms] for each movement trial. The high-dimensional time series for each channel × frequency $x_{w,t,c}$ for [$t_{-1000}$, $t_{-950}$, …$t_0$…$t_{+1000}$] were averaged for each of the 6 movement direction conditions $d = [1…6]$. jPCA was used on these mean timeseries to obtain a set of jPC coefficients that describe a 10-dimensional space to maximize rotational dynamics in these mean ECoG signal trajectories. The original trial segments $x_{w,t,c,d}$ for [$t_{-1000}…t_{+1000}$] were then projected through these jPCs, such that each trial segment is now reduced to a 10D space as $x'_{[jPC1,2],t,d}$ for [$t_{-1000}…t_{+1000}$].

All time points were then labeled as belonging to one of 12 clusters agnostic to condition, using unsupervised hierarchical clustering of all $x'$ time points with Ward's method, which recursively locates clusters that minimize within-cluster distance variance in the dimensionally-reduced Euclidean space. This method generated a list of observed states $s_{t,d}$ for t = [−1000ms … +1000ms], where all observed s are categorical class labels, $s \in [1…12]$, and $d$ is one of six directions of movement.

The pre-movement cluster-time-series were then isolated and used as the sole training data to construct the HMM. Cluster labels immediately preceding movement $s_{t,d}$ for t = [−500ms…0ms] were grouped by movement direction condition, and used as observed sequences to train six hidden Markov models $H_d$, one for each force movement direction with $d = [1…6]$. The Baum-Welch algorithm was used with 12 observed states and 8 hidden states to train each $H_d$. Direction predictions were made with leave-one-out validation, where each HMM generated from training $s_{t,d}$ was used to calculate the posterior state probability of observing the sequences in the testing $s_{t,d}$. The direction associated with the $H_d$ giving the highest posterior probability (smallest negative log-probability) was used as the classified direction.

*Regularized discriminant analysis:* The normalized, binned time series were split into 1-second chunks aligned to the initiation of movement segments, reflecting a time period of [−1000ms…0ms] for each trial, with the start corresponding to 1 second before movement initiation. Each feature $x_{w,t,c}$ at wavelet scale $w$, timepoint $t$, and channel $c$ is used as a feature for the training of a multiclass regularized discriminant analysis model with 10-fold validation, where the posterior probability that observation $x$ belongs to class $k$ follows

$$\hat{P}(k|x) = \frac{P(x|k)P(k)}{P(x)} = \frac{\frac{1}{(2\pi|\Sigma_k|)^{\frac{1}{2}}}\exp\left(-\frac{1}{2}(x-\mu_k)^T \Sigma_k^{-1}(x-\mu_k)\right) P(k)}{P(x)} \quad (1)$$

Shrunken centroids regularization as described in [18] attempts to reduce overfitting bias by (a) introducing a modified covariance matrix to stabilize the cross-validated sample covariance $\Sigma$ and (b) applying a threshold to a

modified correlation matrix to prune down the number of useful predictors, as described by

$$\tilde{\Sigma} = (1-\gamma)\Sigma + \gamma D \quad (2)$$

$$(x-\mu_0)^T \tilde{\Sigma}^{-1}(\mu_k - \mu_0) = \left[(x-\mu_0)^T D^{-1/2}\right]\left[\tilde{C}^{-1}D^{-1/2}(\mu_k - \mu_0)\right] \quad (3)$$

where,
$$D = \text{diag}(\hat{X}^T\hat{X}), \quad \tilde{C} = (1-\gamma)C + \gamma I \quad (4)$$

$$\tilde{C}^{-1}D^{-1/2}(\mu_k - \mu_0) \geq \delta \quad (5)$$

This is accomplished by iteratively searching through parameters $\gamma$ and $\delta$, where $\gamma$ modifies the sample covariance matrix $\Sigma$, and $\delta$ thresholds the correlation matrix $C$ and reduces the number of predictors by thresholding them from the posterior probability evaluation [18].

### D. Movement Onset and Force Magnitude Prediction

The normalized binned time series was fit to the magnitude of the force using least absolute shrinkage and selection operator (Lasso) regression [19]. Two-fold validation was used wherein the whole time series of the each half of the recorded movement session $r_{w,c}$, with channel × frequency wavelet coefficients through time was used to fit linear regression coefficients $\beta$ where predicted $\hat{y}_t = \beta_{w,c,t} r_{w,c,t} + \beta_0$. This generates $\beta_{w,c,t}$ as a function of a regularization parameter $\lambda$, which thresholds the number of nonzero $\beta$ coefficients. $\lambda$ was chosen such that the mean-squared-error is the lowest in validation, which resulted in 132 nonzero $\beta$ coefficients, which were used to predict the forces in the other half of the recorded session. The predicted force time series $\hat{y}_t$ were then smoothed using a 1st-order 2 Hz, −10dB Butterworth lowpass filter and numerically differentiated to obtain $d\hat{y}_t/dt$. Peak analysis was used to find the local maxima of $d\hat{y}_t/dt$ using a fixed threshold, and the local maxima points represent potential times of force onset.

## III. RESULTS

### A. Force Direction Classification Using jPCA-RR-HMM

An average HMM classification accuracy of 35.6% was obtained using pre-movement clustered timepoints 500ms prior to the onset of movement. Accuracies were computed using leave-one-out cross-validation between 6 force application directions using jPCA-projected, hierarchically clustered pre-movement ECoG. This performance is signifi-cantly higher than chance of 17%. Figure 1 shows the confusion matrix of classification in the 6 directions, where the main diagonal shows the correct classifications. Five of six directions showed accuracies above chance, with the most accurate direction being isometric movement back toward the torso, which also had very low false-prediction rates.

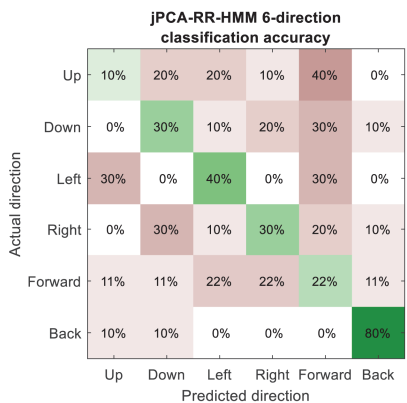

Figure 1. Confusion matrix of the model classification accuracies using jPCA-RR-HMM, in 59 valid isometric force application trials in 6 directions using pre-movement ECoG recordings.

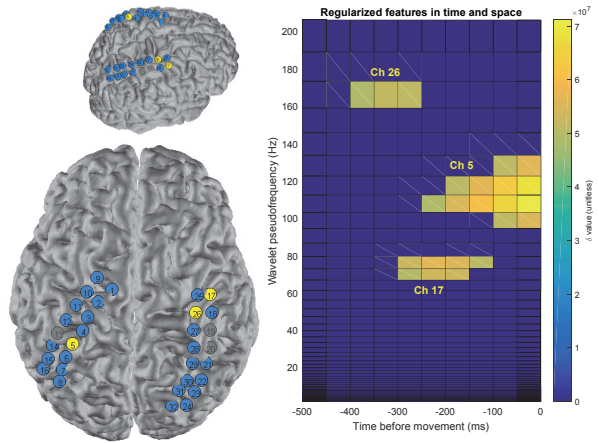

Figure 2. Left: Blue = recorded electrodes; Yellow = recorded electrodes corresponding to regularized features in Right plot; Gray = rejected for artifacts. Right: Time-space plot of salient features from regularization. Each block refers to one feature, most salient (high δ) features in yellow.

### B. Time-Space Localization of Directional Information

In a second analysis, we used regularized discriminant analysis directly on the wavelet features of recorded pre-movement ECoG signals to obtain a cross-validated directional classification accuracy of 27%. While this is not as high as the outcome obtained with the jPCA-RR-HMM analysis, RDA regularization and shrinkage operates on the original feature space, and the remaining 27 features included in computing the posterior class probability must exceed the regularization threshold set by $\delta$ as outlined in equation (5). These remaining wavelet coefficient features demonstrably contain cross-validated directionally-discriminable information, and are localized in time and space. Interestingly, the spatiotemporal distribution of these features as shown in Fig. 2 reveal that the earliest information about movement is present in channel 26 (dorsal premotor cortex), then passes on to channel 17, and finally to channel 5 (primary motor cortex).

### C. Lasso Regression Onset and Force Prediction Results

Cross-validated force predictions agree with overall force magnitude ($R^2 = 0.42$), but a strong predictive relationship between peak force magnitude and ECoG signals magnitude in this task was not found. However, the smoothed predicted force time-series can be numerically derived to obtain an accurate assessment of force onset, predicting onset to within 250ms at a 60% accuracy and within 500ms at an 80% accuracy. The false positive predictive rate was 22%.

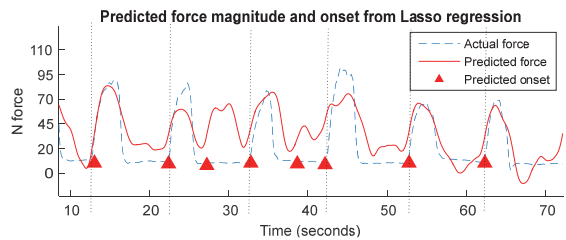

Figure 3. Force and onset timing predictions from Lasso regression. Triangles: predicted force onset; dotted vertical lines: actual onset.

IV. DISCUSSION

A. Direction-Discriminating Dynamics

We showed that the combination of dimensionality reduction and HMM allowed for RR-HMM states to capture complex multidimensional dynamics in a relatively small number of observed states. The performance of this technique suggests that the dynamics between ECoG timepoints during movement preparatory behavior may carry useful task-relevant information. Though the predictive performance of this technique is not sufficient alone to drive brain-computer interfaces, it may be used in combination with techniques such as Kalman filters that can benefit from earlier probability updates to improve the inference of user intent.

B. Spatiotemporal Feature Localization

We found that wavelet-based spectral characteristics of ECoG recordings can discriminate between different directions of isometric arm movements significantly better than chance in a regularized discriminant analysis, despite the overlap between the recruited muscle groups and the isometric task. We find that signals carrying direction-discriminatory information occurred over the ipsilateral premotor cortex up to 400ms before the onset of detected force, in contrast to the signals from the contralateral sensorimotor cortex which showed the strongest discrimination ability starting roughly 100−150msec prior to movement onset (Fig. 2). The timing of the information in ipsilateral changes is particular interesting in potentially offering a technique to screen for neural processes spanning multiple cortical regions, and may give hints to the possible timing of distributed trajectory planning processes.

C. Differences Between Movement Trajectory and Magnitude

While the ECoG signal can be used to both decode direction and force onset in parallel, we found in this dataset that high-performing regression features in predicting force changes do not also perform well in discriminating between directions, potentially suggesting separate usable ECoG signatures for motor execution and planning.

D. Future Considerations

Further work is necessary to determine whether these particular correlates of movement planning generalize across tasks to other types of upper-limb movements, or generalizes to other measures of force application such as non-task muscle contraction. We also wish to examine the possibility of learning effects with repeated practice.

ACKNOWLEDGMENT

We would like to thank Lila Levinson for her outstanding exploratory analyses on this dataset.